\documentclass[%
 aps,
 prx,
 reprint,
 superscriptaddress,
 amsmath,amssymb,
floatfix,
]{revtex4-2}

\usepackage{graphicx}
\graphicspath{{figures/}}

\usepackage{dcolumn}
\usepackage{bm}
\usepackage{braket}
\usepackage[normalem]{ulem}
\usepackage[
range-units = single,
retain-zero-uncertainty = true,
per-mode = symbol,
list-units = bracket,
list-final-separator = {,~}
]{siunitx}
\DeclareSIUnit{\gauss}{G}
\DeclareSIUnit{\quanta}{quanta}

\usepackage{xspace}
\usepackage{isotope}
\newcommand{\calcium}{$\isotope[40]{Ca}^+$\xspace}

\usepackage{xcolor}
\usepackage[colorlinks=true, allcolors=blue]{hyperref}

\usepackage{verbatim}

\usepackage{xstring}
\newcommand*{\aref}[1]{%
	\IfBeginWith{#1}{eq:}{Eq.~\eqref{#1}}{}%
	\IfBeginWith{#1}{fig:}{Fig.~\ref{#1}}{}%
	\IfBeginWith{#1}{tab:}{Table~\ref{#1}}{}%
	\IfBeginWith{#1}{appendix:}{Appendix~\ref{#1}}{}%
	\IfBeginWith{#1}{sec:}{Section~\ref{#1}}{}%
}

\usepackage{subfiles}

\newcommand{\thetitle}{
    Multi-zone trapped-ion qubit control in an integrated photonics QCCD device
}

\newcommand{\affiliationTiqi}{
\affiliation{Institute for Quantum Electronics, ETH Z\"{u}rich, 8093 Z\"{u}rich, Switzerland}
}

\newcommand{\theauthors}{

\author{Carmelo Mordini}
\email{cmordini@phys.ethz.ch}
\thanks{These two authors contributed equally}
\affiliationTiqi
  
\author{Alfredo Ricci Vasquez}
\thanks{These two authors contributed equally}
\affiliationTiqi

\author{Yuto Motohashi}
\affiliationTiqi

\author{Mose M\"uller}
\affiliationTiqi

\author{Maciej Malinowski}
\affiliationTiqi
  
\author{Chi Zhang}
\affiliationTiqi

\author{Karan K. Mehta}
\altaffiliation[Current address: ]{School of Electrical and Computer Engineering, Cornell University, Ithaca, NY 14853, USA}
\affiliationTiqi

\author{Daniel Kienzler}
\affiliationTiqi

\author{Jonathan P. Home}
\email{jhome@ethz.ch}
\affiliationTiqi
\affiliation{Quantum Center, ETH Z\"{u}rich, 8093 Z\"{u}rich, Switzerland}

}

\begin{document}

\title{\thetitle}
\date{\today}
\theauthors

\begin{abstract}
Multiplexed operations and extended coherent control over multiple trapping sites are fundamental requirements for a trapped-ion processor in a large scale architecture. Here we demonstrate these building blocks using a surface-electrode trap with integrated photonic components which are scalable to larger numbers of zones. We implement a Ramsey sequence using the integrated light in two zones, separated by \SI{375}{\micro\meter}, performing transport of the ion from one zone to the other in \SI{200}{\micro\second} between pulses. In order to achieve low motional excitation during transport, we developed techniques to measure and mitigate the effect of the exposed dielectric surfaces used to deliver the integrated light to the ion. We also demonstrate simultaneous control of two ions in separate zones with low optical crosstalk, and use this to perform simultaneous spectroscopy to correlate field noise between the two sites. Our work demonstrates the first transport and coherent multi-zone operations in integrated photonic ion trap systems, forming the basis for further scaling in the trapped-ion QCCD architecture.
\end{abstract}

\keywords{}

\maketitle

\section{Introduction}

Large-scale quantum systems using individual atomic qubits offer promising approaches for quantum information processing \cite{Haffner2008, bruzewicz_trapped-ion_2019, kaufman_quantum_2021} and metrology \cite{RevModPhys.87.637}.  Such applications require pristine control over the atomic systems, implemented via precise and stable delivery of electromagnetic fields. One possible way to scale up the control capabilities while achieving high operational fidelities is to distribute both the qubits and their control fields over multiple spatially distinct zones of a trap device. Each zone contains only a small number of qubits, manipulated via local operations and connected to other zones by physically transporting information carriers, often the atoms themselves, from site to site. This idea has been first proposed and demonstrated with trapped ions, where it is a fundamental component of the Quantum Charge-Coupled Device (QCCD) architecture \cite{kielpinski_architecture_2002, Home2009, Kaushal2020, Pino2021, hilder_fault-tolerant_2022, moses2023race}, as well as more recently for neutral atoms in optical tweezers \cite{beugnon_two-dimensional_2007,bluvstein_quantum_2022}. 

A relevant challenge for scaling such systems is the delivery of the laser light used to coherently control the qubits. For trapped ions, the traditional approach based on free-space optics becomes challenging due to the combined constraints of creating multiple tightly focused beams and the presence of nearby trap electrodes. This complexity increases significantly when multiple wavelengths of light are required in each zone.
An alternative approach is to deliver light using optical waveguides directly integrated into the trap structure. This provides compact routing of light to the zones of interest, where grating couplers allow it to be focused directly on the trapped particles \cite{Mehta2016, bruzewicz_trapped-ion_2019, Mehta2020, Niffenegger2020, Ivory2021_integrated_yb, malinowski_2022_generation, mehta2023_spie_opto}.

\begin{figure}[ht!]
  \includegraphics[width=\columnwidth]{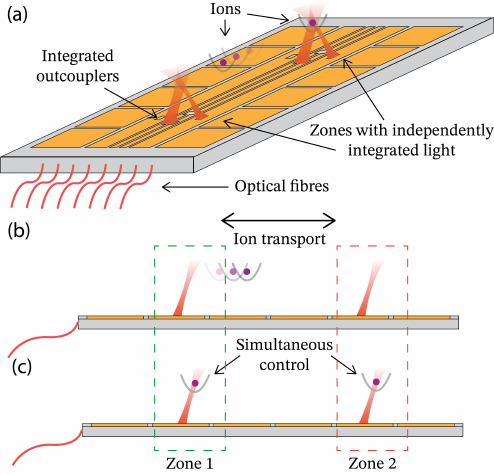}
  \caption{\textbf{Multizone surface-electrode trap with integrated photonics.}
  (a) We demonstrate building blocks of QCCD architecture in a surface-electrode trap with integrated photonic elements. Light is coupled to the trap using optical fibers, and integrated waveguides are used to deliver red and infrared light to two different trapping zones. (b) We perform a Ramsey experiment between the two separate zones using the integrated light. This is enabled by methods to mitigate the effect of photonic structures on ion transport. (c) We simultaneously and independently control two ions in two different zones and correlate spectroscopic measurements across the two sites.}
  \label{fig:cartoon}
\end{figure}

Integrated optics light delivery has thus far primarily been implemented with ions in a single zone with ingredients including delivery of multiple wavelengths \cite{Niffenegger2020} as well as all basic quantum operations including multi-ion quantum logic \cite{Mehta2020}. In parallel with our work, Kwon et al. realized simultaneous addressing of individual ions located in different zones of the same trap \cite{kwon_multi-site_2023}. However, none of these prior works have demonstrated in-sequence transport between zones, in which maintaining low excitation is important.
The use of integrated components in multiple trap zones introduces several complications, as exposed dielectric can affect trap performance through additional heating or charging of surfaces \cite{Mehta2020, Ivory2021_integrated_yb}, which can play a detrimental role in transport routines.

\begin{figure*}[t!]
  \includegraphics[width=\textwidth]{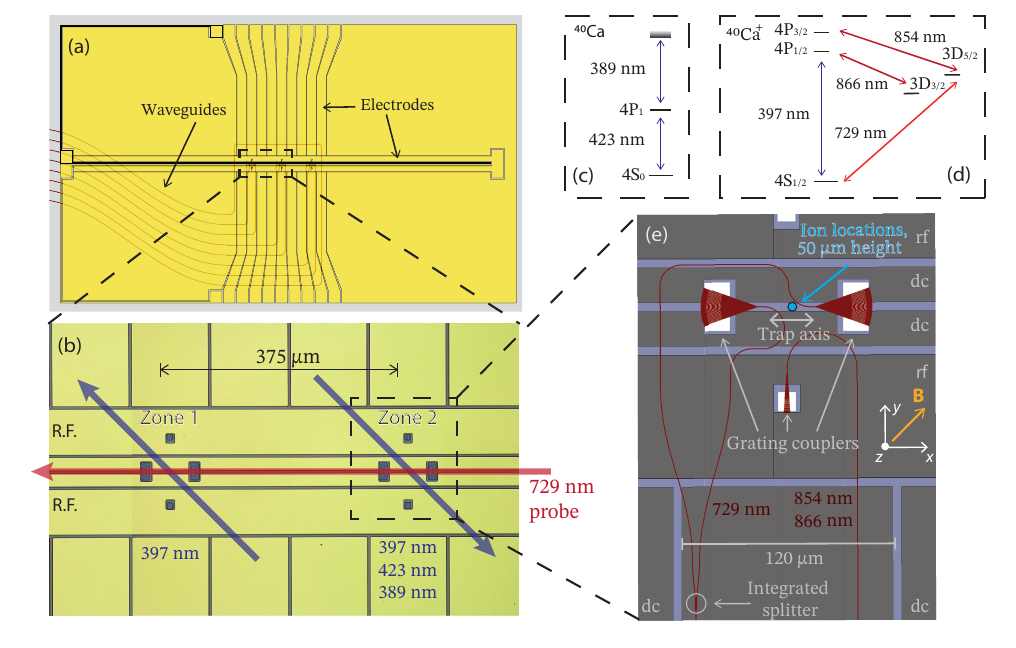}
  \caption{
  \textbf{Details of the trap chip and energy level structure of the ions.}
  With a series of zoomed-in images, we describe the geometry of both the electrical and photonic layers of the trap.
  (a) The large-scale layout shows the relative placement of electrodes and integrated waveguides. The width of the waveguides is exaggerated for better visibility.
  (b) Microscope image of the region of interest for our experiment, showing the two zones where experiments are carried out. Here we highlight the integrated output couplers for each zone and the set of electrodes used to transport and position the ion(s). In overlay, we show a sketch of the free-space laser beams. Near-UV light at \SI{397}{\nano\meter} for cooling and detection is sent to both zones independently, while photoionization beams at \SI{389}{\nano\meter} and \SI{423}{\nano\meter} are sent only to Zone 2. The axial \SI{729}{\nano\meter} beam probes both zones simultaneously. 
  (c-d) Relevant wavelengths for ionizing (c) and controlling the ion (d). 
  (e) The close-up shows the details of the integrated photonic elements in each zone. Two output couplers located on the axis of the trap emit light at \SI{729}{\nano\meter}, while a third one emits light at \num{854} and \SI{866}{\nano\meter} for repumping \cite{ricci_standing-wave_2023}.
  }
  \label{fig:setup}
\end{figure*}

In this work, we demonstrate manipulation and transport of trapped-ion qubits across multiple zones of a surface electrode trap with light delivered using photonics integrated into the trap structure (\aref{fig:cartoon}(a)).
We demonstrate coherence between multiple zones through a distributed Ramsey experiment, which required mapping optical to ground state qubits to avoid sensitivity to the optical phase of the driving beams (\aref{fig:cartoon}(b)). 
With two ions, one in each zone, we demonstrate operation in parallel of multiple regions of the trap, allowing us to characterize operational crosstalk as well as to measure correlations across the device (\aref{fig:cartoon}(c)).
A key element of the work, both for loading and multi-zone operation, is the development and calibration of transport routines that accommodate for differences between the experimental electric field landscape compared to our trap models, primarily due to the presence of the dielectric windows through which light is delivered. 
These ingredients form the primary components for a future scaled multi-zone trapped-ion QCCD device with integrated optical delivery.

The paper is laid out as follows. We first describe the multi-zone trap and its relevant features. We then introduce the development of transport protocols and calibration routines which allow us to mitigate the effects of the exposed windows.
We go on to describe experiments on multi-zone coherent control, in which the ion is transported from zone to zone between operations. Next, we illustrate experiments with parallel operation of multiple zones, and we characterize the crosstalk between them.
Finally, the last sections discuss two technical topics that enabled the experiments described above: independent control of the trap electric fields over multiple zones, and our scheme for scalable light delivery to the different zones of the trap.

\section{Multi-zone trap with integrated photonics}

Our ion trap, sketched in \aref{fig:setup}, is a cryogenic, segmented surface-electrode trap with integrated photonics used to directly deliver light to calcium ions, which can be trapped in multiple zones of the chip. Operation of a single zone of the same device is described in ref. \cite{ricci_standing-wave_2023}.
In the \calcium ion, we use as qubits selected pairs of Zeeman levels within the $4S_{1/2}$ and $3D_{5/2}$ manifolds, addressing them using the quadrupole transition at \SI{729}{\nano\meter}. Additionally, we employ laser light at \SI{854}{\nano\meter} to repump the $3D_{5/2}$ level. Cooling, state preparation, and qubit readout via state-dependent fluorescence are performed using light at \num{397} and \SI{866}{\nano\meter}, while photoionization (PI) beams at \num{423} and \SI{389}{\nano\meter} are used for trap loading. Figure~\ref{fig:setup}(c-d) show the relevant energy levels and wavelengths. 
 
The trap was fabricated in a commercial foundry \cite{worhoff_lionix_2015}, and integrates light at \num{729}, \num{854}, and \SI{866}{\nano\meter} using silicon nitride waveguides.
The waveguides extend to the edge of the chip and are coupled with an efficiency of $\sim \SI{70}{\percent}$ to an attached fiber array \cite{Mehta2020}.
Grating couplers terminate the waveguides for light output, focusing the beams to the position where we trap the ions, at a height of \SI{50}{\micro\meter} above the chip.
The trap has three trapping zones, numbered as in \aref{fig:setup}, out of which we only use Zone 1 and Zone 2. Each of the zones is equipped with one broad-output grating coupler emitting at \num{866} and \SI{854}{\nano\meter}, and two tightly-focusing couplers fed by the same input waveguide, which produce a passively phase-stable standing wave at \SI{729}{\nano\meter} \cite{ricci_standing-wave_2023}.  Voltages applied to the segmented electrodes allow us to shuttle ions along the trap axis $x$ as well as to precisely position them within the intensity pattern of the standing wave in each zone. The layout of the waveguides and electrodes is shown in \aref{fig:setup}(a), (b) and (e).  
While the red and infrared wavelengths are integrated and delivered on-chip, all UV and blue light is delivered in free-space. In this device, two independently controlled beams at \SI{397}{\nano\meter} are sent to Zones 1 and 2, while PI beams at \num{423} and \SI{389}{\nano\meter} are sent to Zone 2, used both as an experimental zone and as the loading zone. 
We also use an additional free-space beam at \SI{729}{\nano\meter} propagating along the trap axis, which we refer to as "axial beam", as a probe for the transport experiments described in this work. The free-space beam geometry is shown in \aref{fig:setup}(b).

In all of our experiments, we begin by Doppler cooling the ion using free-space light at \SI{397}{\nano\meter} as well as integrated \num{866} and \SI{854}{\nano\meter} light. We additionally cool the axial mode of the ion to an average mean phonon occupation of $\bar{n}\sim \SI{1.5}{\quanta}$ using Electromagnetically-Induced Transparency (EIT) cooling \cite{lechner_2016_eit}, with an additional free-space $\sigma$-polarized beam at \SI{397}{\nano\meter} (not shown in \aref{fig:setup}). This beam is then used to initialize the ion in the $\ket{\downarrow} = \ket{4S_{1/2}, m_j=-1/2}$ state via optical pumping. State detection is performed using \SI{397}{\nano\meter} and \SI{866}{\nano\meter} fluorescence. For the experiments described in \aref{sec:parallel}, we initialize the ion in Zone 2 to the $\ket{\downarrow}$ state using integrated light at $\SI{729}{\nano\meter}$ and $\SI{854}{\nano\meter}$.

\section{Stray charge compensation and waveform diagnosis}
\label{sec:charge_compensation}

\begin{figure}
  \includegraphics[width=\columnwidth]{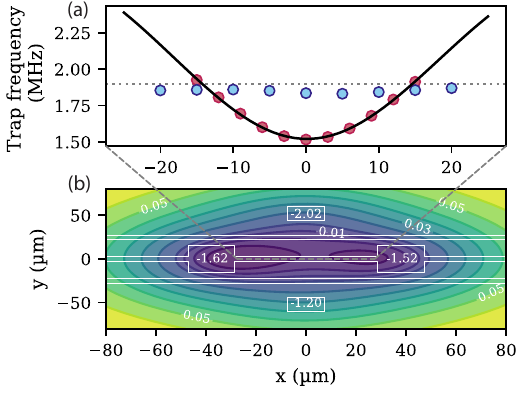}
  \caption{\textbf{Direct measurement of stray potential curvature in Zone 1.}
  We use a trapped ion as a direct probe of the electrical potential above the trap, and use the measurements to model and compensate the discrepancy with the electrical model of the trap.
  (a) Measurement of the local axial trap frequency by sideband spectroscopy using the \SI{729}{\nm} axial beam, as a function of the ion position in Zone 1. The red points show the measured trap frequency without any compensation. The black solid line is a fit of Eq.~\eqref{eq:local_trap_frequency} to these data, yielding the voltage set $\mathcal{V}_1$. The gray, dashed line marks the target trap frequency of \SI{1.9}{\mega\hertz}. The blue data points show the final frequency measurement, where using the voltage set $\mathcal{V}_2$ we achieve a uniform trap frequency. Error bars are smaller than the data points, as the average relative error on the frequency measurements is around \SI{0.5}{\percent}.
  (b) Map of the potential generated by the window electrodes with voltages $\mathcal{V}_2$. The overlying sketch shows the geometry of the coupler cutouts and their voltages, in volts. Labels on the contour lines report the value of the potential in electronvolts. The gray dashed line indicates the span of the $x$ axis covered by the measurements in (a).}
  \label{fig:charge_compensation}
\end{figure}

A crucial step in the experiments described in \aref{sec:multizoneRamsey} is the transport of ions between the two locations. In our trap, the grating couplers face the surface through cutouts in the electrodes, creating windows of exposed dielectric visible in \aref{fig:setup}(b). In these regions, as well as in the gaps between the electrodes, the electrical potential is not controlled by DC voltage sources and hence they are obvious candidates for the creation of undesired potentials due to localized stray charges. In our experiments, a particular challenge is that the presence of those charged regions affects the performance of transport operations, for which we devise strategies to quantify and mitigate their effect.

To transport the ion, we execute a pre-programmed sequence of time-dependent voltages, or waveform, on the trap DC electrodes. The waveform is synthesized by solving a quadratic optimization problem. Starting from a model of the potential generated by each electrode when a unit voltage is applied, the program calculates the voltages that best reproduce the total potential required at each timestep, including a set of physical objectives such as a predefined transport trajectory or a constraint on the maximum voltage output.
The electrode potentials are calculated analytically in the gapless approximation \cite{House2008}, with formulas synthesized for an approximation to our electrode geometry using the Mathematica package \emph{SurfacePattern} \cite{surfacepattern}. In this manuscript we refer to the formulas describing the potentials as the \textit{trap model}. Our code for the generation and analysis of the waveforms has been released as an open-source Python package \cite{Mordini_pytrans}. Further details on the hardware setup are described in \aref{appendix:transport}.

The transport sequence for these experiments moves the ion between Zones 1 and 2 along a heuristic sigmoid trajectory \cite{negnevitski_thesis_2018}. The speed is limited by low-pass filtering of the voltages applied to the trap, implemented with two sets of filters with a combined cutoff of \SI{30}{\kilo\hertz}. We aim for the transport to be as fast as our filter electronics allow while maintaining a constant axial frequency of \SI{1.9}{\mega\hertz} along the trajectory.

We first quantify the effect of directly using sequences based on the modelling by performing qubit rotations after transport. We prepare the ion in Zone 1, then transport it to Zone 2 in \SI{200}{\micro\second} where we drive carrier Rabi oscillations using the \SI{729}{\nm} integrated beam. The ion is then brought back to Zone 1 for detection by running the same waveform in reverse. We measure the average motional excitation from the decay of the Rabi oscillation signal, with a model accounting for coherent and incoherent excitation --- the fitting procedure is described in \aref{appendix:nbar_carrier}.
In initial experiments without compensating for the presence of the windows, we quantify the total excitation delivered by the transport as a coherent component with average Fock number $|\alpha|^2 \sim \SI{58}{\quanta}$, and an incoherent component of $\bar n \sim \SI{25}{\quanta}$, which significantly reduces the quality of even relatively short carrier oscillations required for gates.
We relate this amount of excitation to strong variations in the trap frequency that the ion might be subject to during uncompensated transport. On one hand, the ion can experience increased heating rate if the trap frequency decreases for an appreciable fraction of the waveform time \cite{Brownutt2015_heating_review}. For reference, we report a measured heating rate of between 2 and \SI{3}{\quanta\per\milli\second} for the axial motional mode at a frequency of \SI{1.9}{\mega\hertz}, and we highlight that separate measurements of the heating rates with the ion directly above the windows were not observably different from those measured in other regions of the trap axis, where no windows are present.
On the other hand, an increased trap frequency could lead to crossing between the axial and one of the hot radial modes, resulting in an excitation transfer \cite{Lancellotti2016_crossing}.
In either case, a compensation strategy that stabilizes the value of the trap frequency is beneficial to improve transport.

The trap model assumes the window surfaces to be grounded electrodes, neglecting the effect of the underlying dielectric. To mitigate the effect of the model discrepancy in the region of the windows, we characterized it and built a heuristic model that describes it with additional potentials associated to the windows --- whether arising due to approximations in our trap model or from charge build-up due to laser light --- allowing compensation in the waveform design. In the improved model, we replace the grounded windows with fictitious electrodes, with the same shape as the cutouts, loaded with a set of effective ``coupler voltages'' $\mathcal V = \{V_j\}$. Here, the index $j$ labels the four windows present in each zone in our trap geometry.

Using the ion as a local probe, we determine $\mathcal{V}$ from a spatially resolved measurement of the stray curvature. This is done by positioning the ion along the trap axis in a potential which, according to the model with grounded windows, produces a target trapping frequency $\omega_0 = 2\pi \times \SI{1.9}{\mega\hertz}$ independent of position. 
We attribute deviation of the measured frequencies from this value to the stray curvature generated by the windows, parametrized as $\mathcal{C}(x) = \sum V_j \partial_x^2 \phi_j(x)$, where $\phi_j$ is the unit potential generated by the window electrode $j$. This results in a spatially dependent trap frequency
\begin{equation}
    \label{eq:local_trap_frequency}
    \omega(x) = \sqrt{\omega_0^2 + \frac{q}{m} \mathcal{C}(x)}.
\end{equation}
We measure the actual trapping frequency with sideband spectroscopy using the axial beam, and fit it with \aref{eq:local_trap_frequency} to extract the values of the coupler voltages.
Results of performing this calibration in Zone 1 are shown in \aref{fig:charge_compensation}(a). Red data points show the measurements, while the black solid line shows the fit from which we obtain the voltage set
$\mathcal{V}_1 = \{\numlist{-2.21; -2.14; -2.12; -2.12}\}$~\unit{\volt}, for the left, right, top, and bottom window respectively. While fitting, we constrain the voltages of the top and bottom windows to be the same, since we can only probe the curvature profile along the trap axis.
We then replace the grounded windows with $\mathcal{V}_1$ in the trap model, and synthesize a new voltage set targeting again a homogeneous frequency of value $\omega_0$. We repeat the measurement, and find that we still measure a non-uniform frequency profile, but closer to the expected one. We further update the coupler voltages and iterate the procedure until we measure a uniform frequency close to the target value. After the last iteration we obtain the values $\mathcal{V}_2 = \{\numlist{-1.62; -1.52; -2.03; -1.20}\}$~\unit{\volt}, leading to the trap frequency measurements shown in blue.
In the last round of calibration, we also introduce a difference between the modeled top and bottom window electrode voltages to account for the observed micromotion compensation fields required in the $y$ direction.
Figure~\ref{fig:charge_compensation}(b) shows this final voltage set on the trap layout and a map of the stray potential generated by the window electrodes in a plane parallel to the electrodes at the height of the ion. Although this method could be straightforwardly carried out in Zone 1, in Zone 2 (the loading zone) the discrepancies between model and experiment were larger, which led to problems with ion loss during implementation. Compensation in Zone 2 was therefore performed using transport, as described in \aref{sec:doppler}.

Our heuristic model allows us to compensate for the discrepancy between the predicted and the actual potential, but it does not provide information regarding the source of such discrepancies. 
The first possible source is the electrical response of the other layers of the trap chip underneath the electrodes. This is normally neglected by the gapless approximation used in our analytical model, which breaks down at the windows, whose size is not small compared to that of the nearby electrodes \cite{Schmied_2010_gaps}.
This results in a net charge on the windows induced by the field created by the electrodes, which in the heuristic model would be described by an effective coupler voltage correlated to the ones applied to the surrounding DC electrodes.
We quantify this effect by comparing the prediction of the analytical trap model with the result of a 3D finite element method (FEM) simulation, that calculates the electric potential on the trap surface accounting for all the conducting and dielectric layers of the trap and their geometry. In \aref{appendix:fem_model} we describe in detail the trap layer stack and the simulation and show the obtained results.
From the comparison, we observe that the dielectric response effect is non-negligible, but it is mitigated by the presence of the ground plane, which lies below the electrodes at a distance much smaller than the size of the coupler windows and thus provides an effective grounding for the above dielectric.

We find that this effect cannot account alone for the measured discrepancy. A rough estimate based on the FEM simulation is that the dielectric contributes to around \SI{30}{\percent} of the effective voltage that we attribute to the window. We think that the remaining part can describe charges of other nature, and there are indications that photoinduced charging is a relevant source. We find that the excess voltage is negative, in agreement with negative charging due to the accumulation of photoelectrons on the dielectric substrate \cite{Harlander2010_charging, lee2023photoinduced}. Furthermore, we observe that the loading zone showed a stronger discrepancy with the analytic model, which we correlate with the presence of the PI beams.

\section{Transport diagnosis via Doppler velocimetry}
\label{sec:doppler}

\begin{figure*}
  \includegraphics[width=\textwidth]{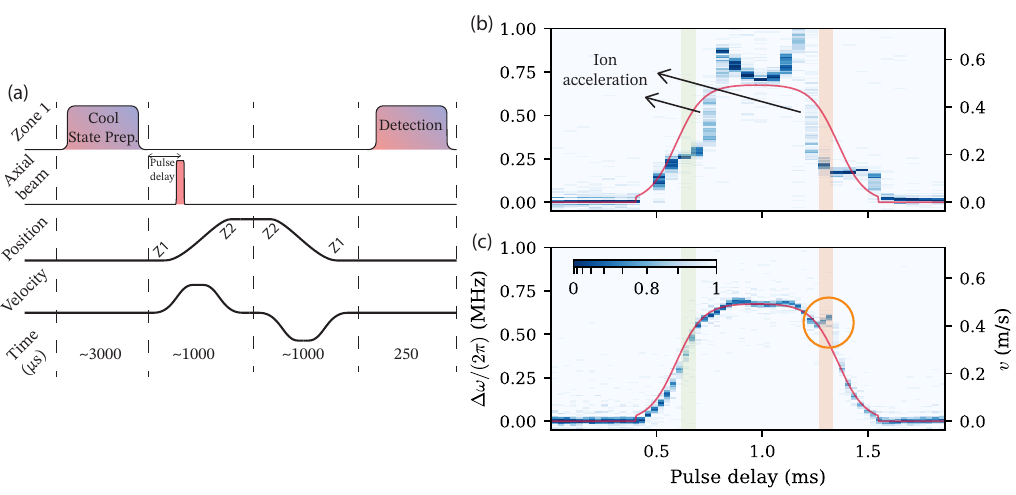}
  \caption{\textbf{Doppler velocimetry.}
  (a) Experimental sequence to measure the velocity profile. The ion is initially cooled and prepared in Zone 1 (Z1), and then adiabatically transported to Zone 2 (Z2) in \SI{1}{\milli\second}. A \SI{30}{\micro\second} \SI{729}{\nano\meter} pulse is applied with a variable time delay during transport using the free-space axial beam. Finally, the ion is brought back to Zone 1 for state detection.
  (b-c) In-flight spectroscopic measurement of the transport velocity of a single ion, probing the Doppler shift induced by transport with a free-space \SI{729}{\nano\meter} beam propagating parallel to the trap axis, before (b) and after (c) compensation of the grating couplers. The colorbar shows the population in the $\ket{\downarrow}$ state. $\Delta\omega$ is the frequency offset of the beam to the qubit frequency of the ion at rest. The scale on the right axis measures the particle velocity calculated from the first-order Doppler shift. The green and orange regions show the times when the ion would be on top of the outcouplers, if it followed the ideal trajectory.}
  \label{fig:doppler}
\end{figure*}

In order to quantify the effectiveness of our stray field compensation method, we directly characterize transport using Doppler velocimetry. Using the axial beam, we perform spectroscopy of the carrier transition at various points while the ion is in motion. From this, we infer the velocity of the ion $\pmb{v}$ from the Doppler shift of the resonance, $\Delta \omega = - \pmb{k} \cdot \pmb{v}$, where $\pmb{k}$ is the beam wavevector.
Figure~\ref{fig:doppler}(a) shows the pulse sequence used in this experiment.
Similar methods have previously been applied for direct probing of the ion position \cite{Clark2023_deshelving}, for the characterization of time-varying laser Hamiltonians \cite{de_clercq_estimation_2016}, and for the implementation of quantum gates \cite{de_clerq_2016_transport_gate, Tinkey2022_transport_entangling_gate}.

With this method, we characterize the transport of one ion from Zone 1 to Zone 2 using a waveform with a total duration of about \SI{1}{\milli\second}, before and after including the coupler voltages in the model.
The spectroscopy pulse has a duration of \SI{30}{\micro\second}, limited by the available laser power in the axial beam. Because of this, in these experiments we use a slower waveform to keep the probe time short compared to the acceleration timescale $|\pmb{v}| / (dv/dt)$, allowing us to measure quasi-instantaneously the velocity profile.
Figure~\ref{fig:doppler}(b) shows the velocity profile measured before introducing compensation. We observe deviations from the expected trajectory (red solid line) and notice that the effect is particularly strong when the ion is moving near the couplers, as it experiences an uncontrolled acceleration caused by a transient potential. This can be observed as a spread in the spectroscopy signal, as the Doppler-shifted resonance frequency sweeps over different values within the time of the probe pulse.
We then include compensation in the generation of the transport waveform, using for both zones the voltage set $\mathcal{V}_2$ which we measured in Zone 1. This is effective in removing the excitation observed at the beginning of transport close to Zone 1, but is not sufficient to avoid it in Zone 2. We then adjust the voltages in Zone 2 to suppress the residual excitation measured in the velocity profile. We do this by synthesizing multiple waveforms with different coupler voltages in Zone 2, and choosing the set where the velocity profile is closest to the intended trajectory. The number of possible voltage sets tested in this way is limited by the long time needed to complete the Doppler velocimetry measurement, and therefore the method described in \aref{sec:charge_compensation} is generally preferable.
The final compensation voltages for Zone 2 are $\{\numlist{-3.14; -3.14; -3.62; -2.82}\}$~\unit{\volt}, which results in the trajectory shown in \aref{fig:doppler}(c). There remains an observable discrepancy between the expected trajectory and the data near the couplers in Zone 2, which is the loading zone.
We think that compensation there is made harder by the frequent ion reloading, which affects the charging of the windows.

We use the refined set of corrections to generate faster waveforms, and optimize the waveform time using Rabi oscillations after transport. We find the optimal waveform time to be around \SI{200}{\micro\second}. Slower waveforms result in unwanted motional excitation due to the high heating rates of the trap. Faster waveforms resulted in a reduced quality of Rabi oscillations, which we attribute to waveform distortion due to the in-cryo filters, as described in \aref{appendix:transport}. The waveforms with optimal time are used in the subsequent experiments described below.

We proceed to measure the residual motional excitation along the transport trajectory after including the compensation. We do this by transporting the ion from Zone 1 to Zone 2 and stopping the waveform at different positions along the trap axis. We then perform Rabi oscillations with the axial beam and bring the ion back to Zone 1 for detection.
We measure the motional excitation of the ion fitting the Rabi oscillations with the method described in \aref{appendix:nbar_carrier}. A first analysis that used the mixed thermal/coherent excitation model resulted in an almost negligible incoherent component in the motion excitation. Therefore, we proceed assuming that the ion is in a coherent state after transport, and we measure the evolution of the coherent excitation through the transport sequence.

\begin{figure}
  \includegraphics[width=\columnwidth]{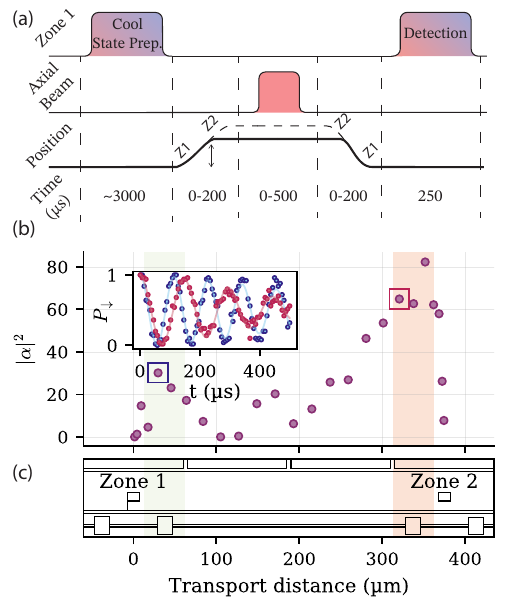}
  \caption{\textbf{Mid-transport measurement of motional excitation after coupler compensation.}
  (a) Experimental sequence to measure the motional excitation along transport. The ion is initially cooled and prepared in Zone 1 (Z1), and then transported to an intermediate position between Z1 and Zone 2 (Z2) by truncating the waveform which includes the coupler compensation.
  We perform Rabi oscillations in that location using the axial beam, and finally, we bring the ion back to Zone 1 for state detection.
  (b) Measured motional excitation as a function of transport distance. The inset shows two sample Rabi flops, corresponding to the data points squared with the same color. (c) A sketch of the trap layout between the two zones.}
  \label{fig:excitation-position}
\end{figure}

Figure~\ref{fig:excitation-position}(a) illustrates the pulse sequence used for this experiment, while (b) shows the measured $|\alpha|^2$ as a function of the final position in the transport. For reference, we draw in (c) a sketch of the relevant region of the trap.
We observe a transient excitation when the ion passes above the windows, reaching up to $|\alpha|^2 \sim \SI{80}{\quanta}$ while approaching Zone 2. We interpret this as the effect of residual miscompensation of the potential near the couplers, as it correlates with the discrepancy in the transport velocity trajectory observed in \aref{fig:doppler}(c).
The ion gets successively de-excited further in the waveform, showing that the residual excitation is fully coherent. This is consistent with the fact that the Rabi oscillations are well-fitted without the needed to include an incoherent motional excitation in the model. When the transport is executed to the end, we observe a final $|\alpha|^2 \sim \SI{8}{\quanta}$.
The reduction in motional excitation from $|\alpha|^2 \sim 58$ to \SI{8}{\quanta}, and $\bar n \sim \SI{25}{\quanta}$ to a negligible amount, for the coherent and incoherent fraction respectively, confirms the effectiveness of our heuristic compensation method. Supplementing the trap model with a description of the potential coming from the coupler regions allows us to synthesize the transport waveform with reduced motional excitation, enabling multizone operations in the integrated photonics architecture.

\section{Multi-zone coherent operations}
\label{sec:multizoneRamsey}
\begin{figure*}
  \includegraphics[width=\textwidth]{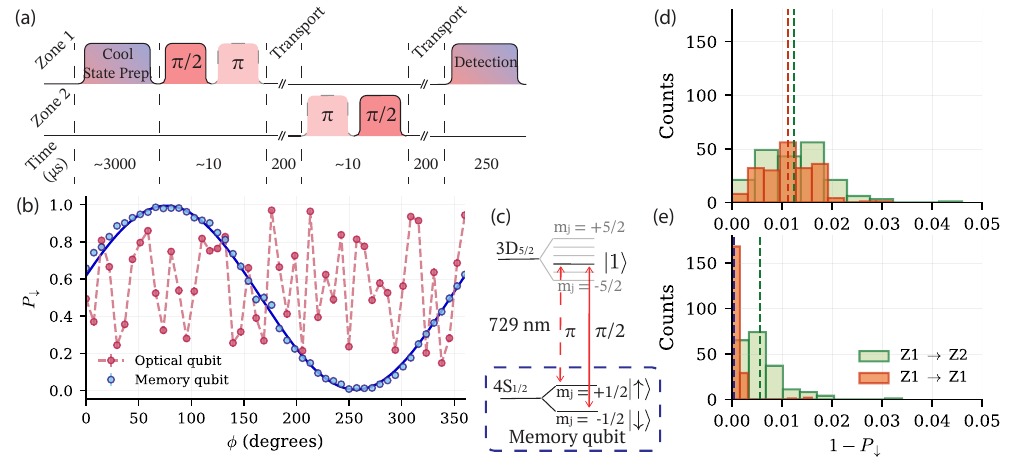}
  \caption{\textbf{Ramsey experiment between zones.}
  (a) Experimental sequence. The ion is initially cooled and prepared in Zone 1. A $\pi/2$ pulse using the integrated \SI{729}{\nano\meter} light creates a superposition state in the optical qubit. The ion is then transported to Zone 2 in \SI{200}{\micro\second} where the second Ramsey pulse is applied using integrated light. Optional $\pi$-pulses are applied to map the optical qubit into the memory qubit. Finally, the ion is transported back to Zone 1 for detection. 
  (b) Population on the $\ket{\downarrow}$ state as a function of the phase of the second Ramsey pulse for the optical and memory qubit. For the memory qubit, we retrieve a Ramsey contrast of \num{0.989 \pm 0.005}.
  (c) Energy levels showing the Zeeman structure of the \SI{729}{\nano\meter} transition. The optical qubit is created using a ground state and a metastable state, while the memory qubit is created using the two Zeeman sublevels of the ground state. A $\pi$-pulse is used to map the population from the metastable state to the ground state.
  (d-e) Histogram showing the state infidelities when applying two consecutive $\pi$-pulses, where in one case (orange) the two pulses are applied in Zone 1 and in the second case (green) the second $\pi$-pulse is applied after transport in Zone 2. The mean infidelities are plotted in dashed lines. Panel (e) shows the infidelities of the final state when using BB1 composite $\pi$-pulses.
  }
  \label{fig:multizone_ramsey}
\end{figure*}

Using the transport developed in section \ref{sec:charge_compensation}, we now proceed to demonstrate multi-zone coherent operations in our integrated device, by performing a  Ramsey experiment across the two trap zones separated by \SI{375}{\micro\meter}.

In a first experiment, we implement the sequence on an optical qubit, defined by the states $\ket{\downarrow} = \ket{4S_{1/2}, m_j=-1/2}$ and $\ket{1} = \ket{3D_{5/2}, m_j=-1/2}$. The ion is initialized in Zone 1 to the $\ket{\downarrow}$ state. We then apply a $\pi/2$-rotation using a laser pulse at \SI{729}{\nano\meter}, which prepares an equal superposition of the states $\ket\downarrow$ and $\ket 1$.
We subsequently transport the ion in a total time of \SI{200}{\micro\second} to Zone 2, where we complete the Ramsey sequence with a second $\pi/2$-pulse. Both pulses are implemented using integrated light, delivered to each zone by separate optical fibers and controlled by individual fiber Acousto-Optic Modulators (AOMs). Finally, we transport the ion back to Zone 1 where we perform detection, measuring the population in the $\ket{\downarrow}$ state via state-dependent fluorescence.
Figure~\ref{fig:multizone_ramsey}(b) shows the population as a function of the phase of the second $\pi/2$-pulse $\phi$. The red points (connected by a dashed line) show the results of experiments performed using the optical qubit. The expected sinusoidal signal is completely scrambled, but the presence of data points close to 1 indicates that the Ramsey sequence can still be closed with high contrast.
The data highlights the presence of slow fluctuations in the relative optical phase between the two laser beams that are used to manipulate the qubit superposition in each of the zones, originated by the optical fibers bringing light to the chip. These effectively randomize the value of $\phi$ for each data point.

The light is generated for both zones by the same narrow-linewidth laser source and is delivered to the experiment with a fully fiberized optical system.
After amplification, the light is split in two using a 50:50 fiber splitter, after which each branch is sent through a separate fiber AOM for pulsing and frequency control, and finally to the experiment via an in-cryo optical fiber directly coupled to the chip.
The extended use of fiber optics, albeit enabling scalability and greatly easing the maintenance of the optical setup, results in fluctuations in the relative phase between the different optical paths, originating from thermal or mechanical stress on the optical fibers.
For small-scale setups, these can be passively mitigated with careful thermal and mechanical anchoring of the fiber setup, by stabilizing the power duty cycle of the driving AOMs \cite{flannery2022_crosstalk}, or by active phase noise cancellation \cite{Ma94_fnc, Noe2023_fnc_earthquakes}. However, as the size of an experiment scales up, implementing these methods to keep track of the relative phases between multiple pairs of beams can quickly become unfeasible.

To mitigate the phase fluctuations, we implement a hybrid encoding scheme by mapping the qubit state from the optical qubit to the ground state manifold using a \SI{729}{\nano\meter} $\pi$-pulse prior to transport, with the qubit now defined by $\ket{\downarrow} = \ket{4S_{1/2}, m_j=-1/2}$ and $\ket{\uparrow} = \ket{4S_{1/2}, m_j=+1/2}$ \cite{chi_zhang_thesis_2022}. In Zone 2, a second $\pi$-pulse reverts the ion state to the optical qubit before the final pulse of the Ramsey sequence.
For detection, we again transport the ion back to Zone 1. The complete sequence is illustrated in \aref{fig:multizone_ramsey}(a). This scheme has the advantage that it does not require phase coherence between the two sets of optical pulses executed in the different zones, but only within pulses in the same zone, which is easier to maintain.
The blue data points in \aref{fig:multizone_ramsey}(b) show the expected oscillation associated with the Ramsey phase scan. In \aref{appendix:hybrid_qubit_math} we discuss the hybrid scheme in more depth and show its insensitivity to the laser phase.

The contrast of the observed Ramsey fringe is \num{0.989 \pm 0.005}.
Possible causes of infidelity are spin decoherence, area errors in the Ramsey pulses, and motional effects.
An independent measurement of the spin coherence of the memory qubit in a single zone, probed directly with a radiofrequency drive, resulted in a Gaussian decay profile with a coherence time of $\sim\SI{12}{\milli\second}$ \cite{malinowski_thesis_2021}. The expected contrast after the $\SI{200}{\micro\second}$ of transport time would be \num{0.9997}, much higher than observed, for which we conclude that spin decoherence has a negligible effect.

We benchmark the quality of the $\pi$-pulses by applying two consecutive pulses in the optical qubit, which ideally would recover the $\ket{\downarrow}$ state perfectly. The histograms in \aref{fig:multizone_ramsey}(d) show the sampled probability distributions for measuring the ion in the state $\ket\downarrow$ for two different sequences: one where the ion stays in Zone 1 (orange), and one where the second pulse happens in Zone 2 after transport (green). We measure this probability by averaging the number of events where the ion was detected in the $\ket{\downarrow}$ state over 500 shots.
We repeat this procedure $N = 200$ times and find the mean fidelities $F = \frac{1}{N}\sum_{i=1}^N P(\ket\downarrow)_{i}$  to be \SI{98.88 \pm 0.05}{\percent} when both pulses are applied in the same zone, and \SI{98.76 \pm 0.03}{\percent} when the second pulse happens after transport.
We can improve these fidelities by using composite BB1 $\pi$-pulses (\aref{fig:multizone_ramsey}(e)) which are robust against pulse area errors \cite{wimperis_broadband_1994, mount_error_2015}.
Here we measure the average fidelities to be \SI{99.95 \pm 0.01}{\percent} and \SI{99.44 \pm 0.03}{\percent} when the pulses are performed in a single zone and in different zones, respectively. The error bars in the fidelities are computed as $\sigma_p/\sqrt{N}$ with $\sigma_p$ the standard deviation of $P(\ket\downarrow)$ in the 200 repetitions.
The improvement obtained using composite pulses suggests that Rabi frequency fluctuations are among the limiting factors for the original loss in fidelity. These fluctuations cannot be completely explained by taking into account the ion thermal state, as calculating the expected fidelity for our ion temperature, and taking into account also the radial modes, we would expect a fidelity of \SI{99.44}{\percent}, still higher than what observed. Other sources of fluctuations, such as laser intensity noise, may be the main causes for the observed reduction in fidelity.
The additional decrease in fidelity when transporting can be partially accounted for by the contrast loss induced from a coherent state with $|\alpha|^2 = 8$, which is expected to be $< 10^{-3}$.


\section{Parallel coherent operations in two zones}
\label{sec:parallel}
\begin{figure*}[t]
  \includegraphics[width=\textwidth]{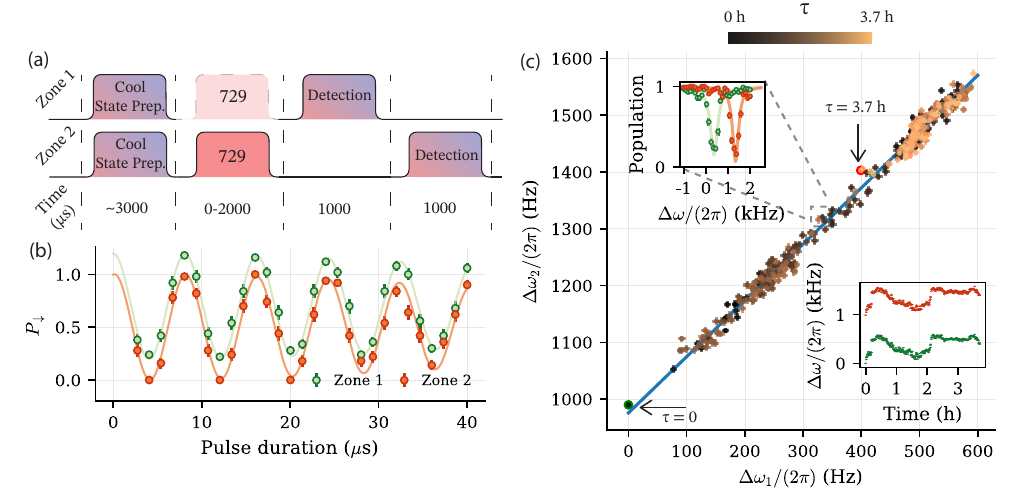}
  \caption{\textbf{Simultaneous control of two ions in different trap zones.}
  (a) Pulse sequence employed for these experiments. Using two sets of \SI{397}{\nano\meter}, \SI{866}{\nano\meter} and \SI{854}{\nano\meter} beams we simultaneously cool and prepare the $\ket{\downarrow}$ state for both ions. Then integrated light at \SI{729}{\nano\meter} is used to control the qubits, followed by two time-resolved detection pulses.
  (b) Simultaneous Rabi oscillations of ions in both zones, with balanced Rabi frequency. The trace for the ion in Zone 1 is vertically offset by 0.2 for better visibility.
  (c) Correlation of simultaneous spectroscopy. The initial ($\tau = 0$) and final ($\tau = 3.7$ h) points are marked with arrows. The color of the points indicates the time when the measurement was performed, where increasing time is depicted with brighter colors. The top inset shows the spectroscopy data for one of the points. The bottom inset shows the time variation of the two qubit frequencies. For all sub-figures and insets, green and orange traces correspond to Zone 1 and 2, respectively.
  }
  \label{fig:simultaneous_flop}
\end{figure*}

We next demonstrate parallel control of the optical qubit $\ket{\downarrow} \rightarrow \ket{1}$ of two \calcium ions each sitting in a different zone of the trap, by driving simultaneous Rabi oscillations on the two qubits. An illustration of the pulse sequence is shown in \aref{fig:simultaneous_flop}(a).
In this configuration, each ion is cooled and prepared in the state $\ket{\downarrow}$ by using an independent set of \SI{397}{\nano\meter} free-space cooling beams, as well as an independent set of integrated \num{866} and \SI{854}{\nano\meter} beams. We use \SI{729}{\nano\meter} light pulsed by two independent single-pass fiber AOMs to control each ion separately, and tune the power of the optical beams to match the Rabi frequencies of both ions. The internal state of both ions is detected by collecting fluorescence onto a single PMT, and we distinguish the two by detecting them at different times.
Results are shown in \aref{fig:simultaneous_flop}(b). From fits, we extract Rabi frequencies of $2\pi\times$\SI{123.2 \pm 0.4}{\kilo\hertz} and $2\pi\times$\SI{124.1 \pm 0.7}{\kilo\hertz} for ions in Zone 1 and 2 respectively.

The ability to control ions in multiple zones in parallel allows us to use simultaneous Rabi spectroscopy to monitor correlations between the qubit frequencies for the two zones. We use a \SI{2}{\milli\second} low-power \SI{729}{\nano\meter} pulse which produces an approximate $\pi$ pulse in each zone.
We select the optical transition $\ket\downarrow \rightarrow \ket{3D_{5/2}, m_j=-1/2}$, with $\Delta m_j = 0$. In our configuration, this transition can be driven in the intensity minimum of the standing wave, minimizing dipole AC Stark shifts \cite{ricci_standing-wave_2023}.
We use Lorentzian fits with central frequencies $\omega_{i}$ with $i\in\{1,2\}$ the zone index, to extract the frequency offset of the transitions relative to a common value. The full-width half-maximum of the fitted Lorentzians is $\sim 2\pi\times\SI{300}{\hertz}$ leading to statistical uncertainties on the frequency measurement of the order of $\sim 2\pi\times\SI{1}{\hertz}$.
We record $\omega_{i}$ for both ions over a total time of \SI{3.7}{\hour} and display the change of the transition frequency in each zone relative to that observed in Zone 1 at time $\tau = 0$ as $\Delta \omega_i = \omega_i - \omega_1(\tau=0)$. Data is shown in \aref{fig:simultaneous_flop}(c).
For each ion, we register frequency drifts of up to $2\pi\times$\SI{592 \pm 6}{\hertz} which might arise either from changes on the magnetic field of up to \SI{1.06 \pm 0.01}{\milli\gauss}, or drifts in the locking cavity of our laser. Decoupling these effects would require probing another transition in the $4S_{1/2} \leftrightarrow 3D_{5/2}$ manifold.
We find an offset frequency between $\omega_1$ and $\omega_2$ of $2\pi\times$\SI{972 \pm 2}{\hertz}, which can be caused by a differential magnetic field of \SI{1.742 \pm 0.004}{\milli\gauss}, corresponding to a magnetic field gradient of \SI{4.98 \pm 0.01}{\gauss/\meter}.
We obtain a correlation coefficient between the $\Delta \omega_1$ and $\Delta \omega_2$ of $R = c_{12}/\sqrt{c_{11}c_{22}} = 0.996$, where $c$ is the covariance matrix. These results are promising for the use of a spectator qubit to monitor and feedback on the magnetic field for other qubits within a multizone architecture \cite{majumder2020_spectator, majumder2020_spectator}. 

\section{Optical crosstalk between trap zones}
\label{sec:crosstalk}
\begin{figure}
  \includegraphics[width=\columnwidth]{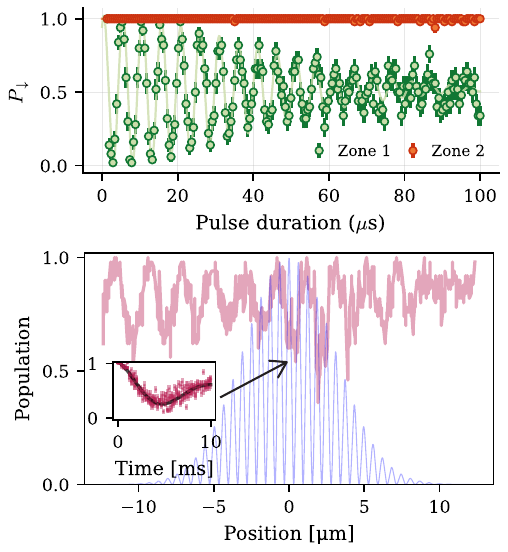}
  \caption{\textbf{Characterization of optical crosstalk between trap zones.}
  (a) Optical crosstalk, measured by performing Rabi oscillations on the ion in Zone 1 and performing state-detection in both ions {(b) Population in the $\ket{\downarrow}$ state of an ion trapped in Zone 1, and driven with light injected in Zone 2. The blue curve in the background shows the simulated axial intensity profile of the light emitted by the output couplers in Zone 1. The inset shows the first Rabi oscillation of an ion located in the position of maximum crosstalk.}
  }
  \label{fig:crosstalk}
\end{figure}

One issue with light addressing both zones is the presence of crosstalk: when an ion in one zone is addressed, the other should not be driven. Figure~\ref{fig:crosstalk} (a) characterizes the optical crosstalk between zones by measuring the population in the $\ket{\downarrow}$ state for both ions while only applying \SI{729}{\nano\meter} light to Zone 1. By performing sinusoidal fits in both datasets, we extract Rabi frequencies of $2\pi\times$\SI{196.3 \pm 0.3}{\kilo\hertz} and $2\pi\times$\SI{0.28 \pm 0.09}{\kilo\hertz} for ions in Zone 1 and 2, respectively, allowing us to bound the Rabi frequency crosstalk in the different zones at \SI{0.14 \pm 0.05}{\percent} in this measurement, or equivalently a power crosstalk in the order of \num{e-6}, in agreement with previous measurements \cite{Mehta2020}.
During this measurement, both ions are positioned at the center of each zone and in the same phase of their respective standing wave pattern, where they experience the maximum Rabi frequency.

To gain insights on the physical origin of the crosstalk, we extended the previous measurement to a broader spatial range. We trap and initialize an ion in Zone 1, send a pulse of 729 light in Zone 2, and measure the ion population in the initial state as a function of its position on the trap axis.
The pulse is calibrated in frequency to resonantly drive Rabi oscillations when directly illuminating the ion,  and has a duration of $\sim\SI{2.5}{\milli\second}$.
We neglect any Stark shifts, given the expected low intensity of the crosstalk pulse.
We scan the ion position over a range of $\pm \SI{12}{\micro\meter}$ above Zone 1, larger than the region illuminated by the integrated grating couplers.
The results of the measurement are shown in \aref{fig:crosstalk}(b), where the curve in the background illustrates the axial intensity profile of the standing wave.
We see a periodicity in the signal, which doesn't match the period of the integrated standing wave. Furthermore, we observe that crosstalk happens also far from the center of the trap zone, where no light is emitted from the couplers of the same zone.
The inset in \aref{fig:crosstalk}(b) shows Rabi oscillations when we locate the ion in the maximum crosstalk position, around the center of the zone. From this measurement we infer a Rabi frequency of $\sim 2\pi \times \SI{0.1}{\kilo\hertz}$, consistent with the previous measurement. 
However, the lowest crosstalk positions exhibits are consistent with suppression of at least an order of magnitude more.
From this we conclude that direct optical coupling from one waveguide to another in the trap substrate is negligible, and that crosstalk arises likely from light scattered from outside the trap. We suspect that the crosstalk might be caused by multiple reflections of the light emitted from the other region of the trap. Further studies would be required to uncover the origin of this effect.

\section{Precise control of electric fields for two-ion operations}
\label{sec:e_field_control}

\begin{figure}
  \includegraphics[width=\columnwidth]{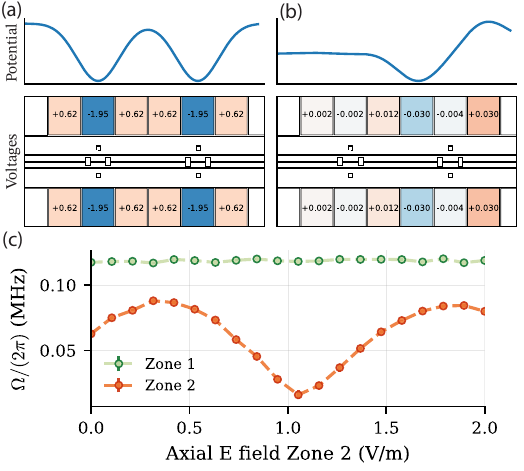}
  \caption{\textbf{Independent control of the ion position in both zones}.
  (a) Trap sketch illustrating the voltage set controlling the axial trapping frequency in the two zones, showing the voltage configuration on the electrodes and the associated axial potential.
  (b) Voltage set controlling the position of the ion in Zone 2. The electrodes of Zone 2 are loaded antisymmetrically to create an axial field, while at the same time the electrodes of Zone 1 are optimized to cancel the spurious field and curvature created by the first ones. A similar voltage set, where the role of the zones is reversed, is used to control the ion position in Zone 1.
  (c) Rabi frequency measurements in the standing wave, used as a proxy for the ion position, performed in both zones while scanning the axial electric field in Zone 2, demonstrating independent control of the position of each ion. Dashed lines are guides to the eye.}
  \label{fig:e_field_control}
\end{figure}

In multizone operations it is necessary to exert independent control over the electric fields in each trap zone for micromotion compensation. Additionally in our trap, due to the use of the standing wave, we require independent, sub-micrometer control of the position of the ion in each zone \cite{ricci_standing-wave_2023}.
We achieve this by finding voltage sets which minimize the influence of the electrodes controlling one trap zone on the other, making the control independent. In our trap, each zone can be controlled by a set of eight electrodes, three on each side of the trap axis dedicated to the zone, plus two along the axis affecting all zones. This offers enough degrees of freedom to create a harmonic potential in each zone, controlling the strength of the axial curvature, the position of the two minima, and the angle of the axes of the radial modes with respect to the trap plane.
We implement multiple sets of voltages, each controlling independently one degree of freedom of the total potential \cite{Allcock_2010_surface_trap}.
One set, whose potential profile and voltages are shown in \aref{fig:e_field_control}(a), creates a positive curvature of \SI{1}{\mega\hertz} strength in both zones.
Two others create an axial electric field in one zone while at the same time using the electrodes of the other zone to null the unwanted fields due to the first --- \aref{fig:e_field_control}(b) shows the set controlling Zone 2. The total multizone potential is generated by applying a linear combination of those voltage sets.

We test the control scheme with a measurement of the Rabi frequency of each ion while scanning the applied electric field in Zone 2. For each field value, we perform Rabi oscillations on both ions and infer their displacement in the standing wave from the Rabi frequency. Results are shown in \aref{fig:e_field_control}(c). We observe that using the optimized voltage set, we can independently displace the ion in Zone 2 by a full period of its standing wave while having a constant Rabi frequency for the ion in Zone 1, suggesting a minimal disturbance of its position \cite{ricci_standing-wave_2023}.

\section{Light delivery for multiple zones}
\label{sec:light_delivery}

\begin{figure}
  \includegraphics[width=\columnwidth]{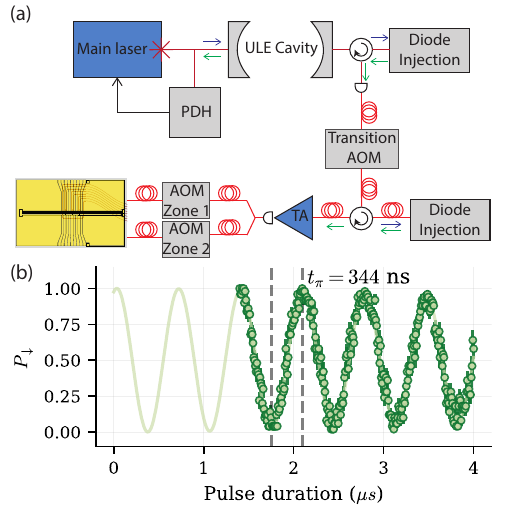}
  \caption{\textbf{Scalable, high-power light delivery to multiple zones.} (a) Simplified optical setup for the qubit laser. Light from the main laser is passed through an ULE cavity, and then further amplified through a first free-space diode injection. A first fiber AOM is used to select the qubit transition, and then light is further amplified with a second injection setup, this time using a pigtailed laser diode and an in-fiber circulator to keep the whole system fiberized. Light is then passed through a tapered amplifier (TA) and then fiber-split into a fiber AOM for each zone before going into the trap waveguides. The white circles with arrows represent optical circulators. (b) Fast Rabi oscillations measured on the $\ket{\downarrow} \rightarrow \ket{3D_{5/2}, m_j = -5/2}$ transition, directing all of the light after the TA into a single zone. The solid line is a fit to the data from where we infer a $\pi$-time of $\SI{344}{\ns}$. Pulses shorter \SI{1.4}{\micro\second} are not possible in the current version of our control system.}
  
  \label{fig:fast-flops}
\end{figure}

Red and infrared light is delivered from a control laser system via fiber AOMs. Light for each of the repumpers is controlled by an independent fiber AOM which is split 50:50 using a fiber splitter and then sent to both zones simultaneously --- thus for these colors individual addressing of repumping in our setup could be achieved by moving ions outside one of the addressed zones.

For the qubit transitions at \SI{729}{\nano\meter}, we gain flexibility from the possibility of driving transitions between the various Zeeman sublevels of the ground and metastable levels.
This is beneficial for hybrid ground-metastable schemes described in the previous sections, as well as for using multiple states as qudits \cite{ringbauer_universal_2022, hrmo_native_2023}.
For this purpose, we implement a scheme allowing us to achieve high powers across the required range of frequencies.
The setup is sketched in \aref{fig:fast-flops}(a). We lock the qubit laser at \SI{729}{\nano\meter} to an Ultra-Low Expansion (ULE) high-finesse cavity using a Pound--Drever--Hall (PDH) scheme. Light transmitted through the cavity is used for the experiment, since this offers filtering of high-frequency noise. This locking scheme results in a laser linewidth below \SI{10}{\hertz} \cite{malinowski_thesis_2021}.
The transmitted light is first injected into a diode and the output is then passed through a continuously running ``transition'' fiber AOM which we use to select which transition is driven in the ion. This AOM has a finite bandwidth and does not produce full power over the relevant range of frequencies to address all Zeeman sublevels (\SI{50}{\mega\hertz} at \SI{6}{\gauss}). We mitigate this effect by using this AOM to seed a second pigtailed diode injection. By continuously running the transition AOM, we can ensure that the second injected diode stays locked and that the output power is constant over the desired frequency range over which the AOM is scanned. Light after the second injection is further amplified using a Tapered Amplifier (TA) before being split using a fiber splitter, after which each branch is controlled by its own fiber AOM, which is used to pulse the beam in the respective zone.
This configuration allows us to input up to \SI{50}{\milli\watt} into each trap zone. Considering the optical losses of about \SI{4}{\decibel} at the input coupling and through the waveguide, and \SI{3}{\decibel} of diffraction efficiency of the grating, we achieve $\sim \SI{10}{\milli\watt}$ of output power at the ion \cite{Mehta2020}. 

We benchmark the maximum coupling rates that can be achieved in our trap by sending the full power output of the TA to Zone 1, passing through its respective fiber AOM. In this way we send up to $\SI{100}{\milli\watt}$ of power into the trap. 
With this configuration, we perform Rabi oscillations in the $\ket{\downarrow} \rightarrow \ket{3D_{5/2}, m_j = -5/2}$ transition. The result is shown in \aref{fig:fast-flops}(b) where we extract a carrier Rabi frequency of $2\pi\times$\SI{1.452(3)}{\mega\hertz}. In order to achieve a high coupling rate simultaneously in multiple zones, we propose to extend the setup by splitting the light after the second injection diode and installing one TA and one control AOM per zone.

\section{Outlook}

Our work demonstrates two key elements for realizing the QCCD architecture with integrated optics, multizone coherent operations as well as parallel independent control over multiple trapping sites. While we obtained a significant improvement using our heuristic model for compensating discrepancies between our electrode model and experiments,  operation fidelities after transport are still limited by the residual motional excitation after transport. This could be further improved by having a dedicated loading zone, away from the experimental zones, which in our experiments could be the third, not yet used trap zone \cite{kwon_multi-site_2023}. At current transport speeds, reducing motional excitation after transport would also require an improvement of heating rates in integrated optics traps \cite{Mehta2020, Niffenegger2020}, or constant re-cooling \cite{Home2009, Pino2021}. Compensation of waveforms may become more critical with non-adiabatic transport \cite{bowler_coherent_2012, walther_controlling_2012, Alonso2016_bang_bang}.

A next step for further scaling in our setup would be to integrate UV light, which will likely require including new waveguide materials in the trap stack \cite{West2019_blue_integrated_photonics, beck2023grating}. In new traps, challenges with exposed dielectric could be mitigated by incorporating conductive coatings on the light emission windows above the grating couplers, for instance using indium tin oxide \cite{Eltony2013_ito_trap, Niffenegger2020, Brown2021_materials}.
Alternatively, the use of an ion species such as barium that requires higher wavelengths \cite{dietrich2009barium, Yum2017_barium, Low2020_barium_qudits}, or schemes for performing all control with longer wavelengths of light \cite{hendricks_doppler_2008, lindenfelser_cooling_2017,zhang_realizing_2017} could help to reduce charging as well as facilitate the integration of all wavelengths.

While the work here observed relative optical phase drifts between zones, we think that these originate in the fibers delivering light to the cryostat, as well as the fiber AOMs used for control. A large part of this could likely be alleviated by classical interferometric approaches, or by maximizing the commonality of the two light paths, for instance using integrated devices \cite{moody_2022_roadmap, Reens2022_spad, knollmann2024integrated, malinowski2023_wiring}. The high level of correlation of the measured qubit frequency drifts at different points in the trap array opens the path to using spectator ions trapped on the same chip as target qubits to perform continuous calibration and stabilization of either the magnetic field or the laser frequency, which can help in the implementation of more complex algorithms \cite{hume_probing_2016, borregaard_efficient_2013, majumder2020_spectator, kim_improved_2023}.

When realized at a large scale, the demonstrated integrated multizone architecture can be beneficial to a range of different applications, both in QCCD quantum computing and for parallel arrays of clocks in metrology.

\begin{acknowledgements}
We acknowledge funding from the Swiss National Science Foundation under Grant No. $200020\_207334$, the National Centre of Competence in Research for Quantum Science and Technology (QSIT),  the EU Quantum Flagship H2020-FETFLAG-2018-03 under Grant Agreement No. 820495 AQTION, and the Intelligence Advanced Research Projects Activity (IARPA), via the U.S. Army Research Office through Grant Number W911NF-16-1-0070 under the LogiQ program. KKM acknowledges support from an ETH Post-doctoral fellowship.

We acknowledge Fabian Schmid and Shreyans Jain for their useful review of the manuscript. 

The experiments were designed, performed, and analyzed by CM, ARV, YM, and MMu. CM and YM developed the software for waveform optimization and analysis. The experiments were performed in a trap designed by KKM, fabricated by LioniX International, and operated in an apparatus with significant contributions from MMa, CZ, and KKM. The paper was written by CM and ARV with input from all authors. The work was supervised by DK and JPH.
\end{acknowledgements}

\begin{appendix}
\section{COMSOL FEM model}
\label{appendix:fem_model}
In the analytic trap model we assume, under the gapless approximation, that the area of the coupler windows is grounded while the other electrodes are set at a controlled voltage. This neglects the response of the underneath dielectric, which develops a surface charge induced on the exposed surface by the field generated by the electrodes, partially accounted for by our heuristic model of coupler voltages.
To verify the extent of this effect, we numerically solve the Laplace equation in 3D with the finite element method (FEM) to calculate the electric potential in the volume of the trap chip, considering the full material stack, and compare the results at the trap surface with the approximate analytic model.
The simulation is performed using COMSOL Multiphysics\textsuperscript\textregistered \cite{comsol}.
\begin{figure}
  \includegraphics[width=\columnwidth]{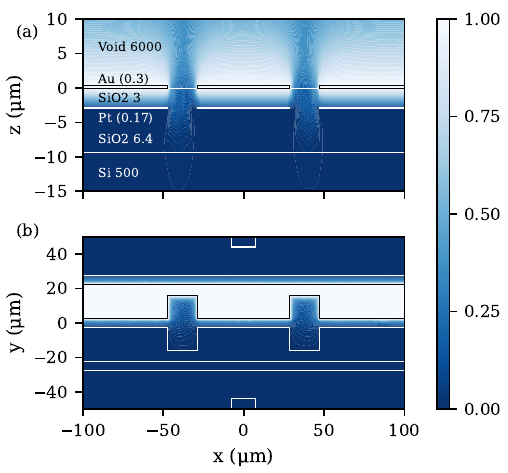}
  \caption{\textbf{FEM simulation of trap-induced charging.}
  Electric potential in the trap volume induced by a potential of \SI{1}{\volt} on the central electrode.
  (a) Slice along a vertical plane passing through a coupler cutout at $y = \SI{10}{\micro\meter}$, the coupler cut plane, and (b) slice along a horizontal plane at $z = 0$, the top surface of the trap. The labels indicate the material of the corresponding layer and the number reports the layer thickness in microns. The thicknesses in parenthesis have been neglected in the simulation. The colorbar on the right side has units of volts, and quantifies the strength of the simulated potential.
  }
  \label{fig:fem_simulation}
\end{figure}

The model includes all the chip layers, illustrated in \aref{fig:fem_simulation}(a). From top to bottom, the stack includes the electrodes (gold, \SI{300}{\nano\meter}), a first dielectric layer (silicon oxide, \SI{3}{\micro\meter}), the ground plane (platinum, \SI{170}{\nano\meter}), another thicker layer of dielectric (silicon oxide, \SI{6.4}{\micro\meter}) embedding the integrated waveguides, and finally a silicon substrate (\SI{500}{\micro\meter}).
Above the trap there is a \SI{6}{\milli\meter} vacuum gap closed by a gold upper boundary, representing a fine gold mesh installed in the experiment to shield the ion trap from the imaging objective present in-vacuum at close distance.
To ease meshing of the 3D volumes while still retaining good simulation accuracy, we neglect the thickness of the electrode and ground plane layers, which are much smaller than the others, and approximate them as planar surfaces. We fully include however their 2D geometry, in particular, we include openings in the ground plane present above the grating outcouplers, in correspondence with the windows in the electrode layer.
As a boundary condition, we set to ground both the gold mesh and the ground plane. We then impose a potential of \SI{1}{\volt} on one of the central split electrodes, which surrounds the coupler windows from one side, and ground all the others.

Figure~\ref{fig:fem_simulation} shows slices of the 3D potential along two cut planes, a vertical plane in (a) parallel to $xz$ and located at $y = \SI{10}{\micro\meter}$, passing through the coupler windows, and the $xy$ plane in (b) at $z = 0$ showing the top surface of the trap. 
The simulation shows how the potential imposed on the electrode penetrates into the dielectric: a \SI{1}{\volt} potential on the electrode produces an average potential of \SI{0.3}{\volt}, mostly localized in the dielectric region surrounded by metal.
This happens within a length scale that is determined by the distance to the ground plane, the closest grounded conductor. The ground plane effectively contributes to lowering the potential measured in the center of the coupler windows, grounding them to a certain extent. We highlight that the effect is present even considering the holes in the ground plane, despite them being placed directly underneath the windows. Due to the presence of the holes, we observe also a potential induced in the second dielectric layer, which justifies the inclusion of the full trap stack in the model.

We validate the FEM model with additional simulations. Setting to ground all the coupler windows and the gap between the electrodes, we reproduce the results of the analytic model in the half-space $z > 0$. Finally, with 2D FEM simulations solving for the potential only along the vertical cut plane, but where we include the thickness of the electrodes and ground plane layers, we verify that neglecting those layers does not impact sensibly the results of the 3D model.

We use this result to estimate the contribution of the dielectric to the total potential on the window. In usual trap operation, both central electrodes are loaded with $\sim \SI{2}{\volt}$, which results in \SI{-0.6}{\volt} distributed over the whole window. This is roughly a \SI{30}{\percent} of \SI{-1.6}{\volt} resulting from our calibration, which represents the totality of the surface charge on the window regardless of its origin.

\section{Laser-phase independence of the hybrid protocol}
\label{appendix:hybrid_qubit_math}
We show how the sequence of operations implemented in our optical/memory qubit hybrid scheme results in effective operations on the memory qubit which do not depend on the phase of the laser field used to manipulate the optical qubit.

The protocol combines rotations of pulse area $\theta$ and phase $\phi$ between different pairs of states chosen among $\mathcal{B} = \{\ket\downarrow, \ket\uparrow, \ket 1\}$. We explicitly write the matrix representation of these operations in the basis $\mathcal{B}$.
A generic rotation on the optical qubit $\{\ket\downarrow, \ket 1\}$ is represented as
\begin{equation}
    U_\text{opt}(\theta, \phi) =
    \begin{pmatrix}
        \cos{\theta/2}  & 0 & ie^{-i\phi}\sin{\theta/2} \\
        0 & 1 & 0 \\
        ie^{i\phi}\sin{\theta/2} &   0 & \cos{\theta/2}
    \end{pmatrix}. 
\end{equation}
Similarly, the $\pi$ rotation mapping between $\ket\uparrow$ and $\ket 1$ is represented as
\begin{equation}
    U_\text{map}(\phi) =
    \begin{pmatrix}
        1 & 0 & 0 \\
        0 & 0 & ie^{-i\phi} \\
        0 & ie^{i\phi} & 0
    \end{pmatrix}.
\end{equation}
To complete the Ramsey sequence, the time spent transporting the ion is modeled as a free evolution of the three-level system, which reads
\begin{equation}
    U_\text{wait}(\tau) =
    \begin{pmatrix}
        1 & 0 & 0 \\
        0 & e^{i \omega_m \tau} & 0 \\
        0 & 0 & e^{i \omega_o \tau}
    \end{pmatrix},
\end{equation}
where $\hbar\omega_m$ is the energy gap in the memory qubit, and $\hbar \omega_o$ is the one in the optical qubit.

Starting from the initial state $\ket\downarrow = (1, 0, 0)^T$, the Ramsey sequence implemented directly on the optical qubit results in the observable
\begin{equation}
\begin{aligned}
    P^o_\downarrow &= |\bra{\downarrow} U_\text{opt}(\pi/2, \phi_{L2} + \phi) U_\text{wait}(\tau) U_\text{opt}(\pi/2, \phi_{L1}) \ket\downarrow|^2 \\
        &= \frac{1 - \cos{(\omega_o \tau - \phi + \phi_{L1} - \phi_{L2})}}{2},
\end{aligned}
\end{equation}
where the laser phases in the two control pulses $\phi_{L1}$ and $\phi_{L2}$ can have randomly different values, and the second control pulse has a deterministic phase offset $\phi$ implemented as a phase shift on the radiofrequency signal driving the acousto-optical modulator used to pulse the beam.
The phase of the Ramsey signal is scrambled by the random value of $\phi_{L1} - \phi_{L2}$, making it impossible to retrieve the contrast.
In the second case, mapping to the memory qubit before and after transport, the sequence reads
\begin{widetext}
\begin{equation}
    P^m_\downarrow = 
    |\bra{\downarrow} U_\text{opt}(\pi/2, \phi_{L2} + \phi) U_\text{map}(\phi_{L2}) U_\text{wait}(\tau) U_\text{map}(\phi_{L1}) U_\text{opt}(\pi/2, \phi_{L1}) \ket\downarrow|^2
    = \frac{1 + \cos{(\omega_m \tau - \phi)}}{2}.
\end{equation}
\end{widetext}
As long as the fluctuations on the optical phase are slow enough such that the control and the mapping pulses share the same value of $\phi_L$, its effect is canceled out in both arms of the Ramsey sequence by the use of this protocol. The Ramsey contrast depends only on the deterministically added phase $\phi$ and can be retrieved with usual analysis methods.

\section{Measurement of transport motional excitation}
\label{appendix:nbar_carrier}

We measure the motional excitation during transport via carrier Rabi oscillations. For an ion in a motional state characterized by a phonon distribution $P(n)$, the Rabi oscillation signal has contributions from all populated Fock states and it reads as
\begin{equation}
    \label{eq:thermal_carrier}
    P_\downarrow = \sum_{n=0}^{\infty} P(n) \frac{1 + e^{-\gamma t} \cos(\Omega_n t)}{2},
\end{equation}
where $\Omega_n = \Omega (1 - \eta^2\,(n + 1/2))$ is the Rabi frequency in the Fock state $n$, valid in the Lamb--Dicke regime, $\Omega$ is the bare carrier Rabi frequency, $\eta$ is the Lamb--Dicke parameter, and $\gamma$ models heuristically any loss of contrast which is not of motional origin, such as laser frequency noise or spin decoherence.
Assuming different motional excitation models, we obtain different Rabi oscillations profiles in \aref{eq:thermal_carrier}, which we use to fit the experimental data and extract the motional excitation parameters.

Assuming a coherent motional state characterized by a displacement $\alpha$, its phonon distribution is a Poissonian
\begin{equation}
\label{eq:populations_coherent}
    P_\text{coh}(n) = e^{-|\alpha|^2}\frac{|\alpha|^{2n}}{n!}.
\end{equation}
Alternatively, to also include incoherent excitation, we use the phonon distribution of a displaced thermal state, resulting from a thermal average of displaced Fock states
\begin{equation}
\label{eq:populations_thermal_coherent}
    \begin{aligned}
        &P_\text{th,coh}(n) = \sum_{k=0}^\infty \frac{\bar{n}^k}{(\bar n + 1)^{k + 1}} |\braket{n | \alpha, k}|^2, \\
        &|\braket{n | \alpha, k}|^2 = e^{-|\alpha|^2}|\alpha|^{2(n + k)}n!k!\, \left| \sum_{l=0}^k \frac{(-1)^l|\alpha|^{-2l}}{l!(n-l)!(k-l)!} \right|^2,
    \end{aligned}
\end{equation}
and characterized by the displacement $\alpha$ and the average thermal excitation $\bar n$.
Even when the sums in \aref{eq:populations_thermal_coherent} are appropriately truncated, it is computationally expensive to directly evaluate the expression even for moderate values of $n$. Therefore, we calculate the distribution $P_\text{th,coh}$ with a numerical thermalization method as described in \cite{Ruster2014_fast_splitting}.
For the measurements reported in \aref{fig:excitation-position}, we begin by separately fitting the first data point of the series, where the ion had not been moved yet, with the model described by \aref{eq:thermal_carrier} and \aref{eq:populations_coherent}. There we set $\alpha = 0$ to extract the base values of the parameters $\Omega$ and $\gamma$. For the rest of the series, we measure the evolution of the coherent excitation by fixing those parameters to the value measured on the first point and fitting $\alpha$ on the remaining data.

\section{Transport waveforms and hardware}
\label{appendix:transport}

We control our DC electrodes using one Sinara Fastino arbitrary waveform generator (AWG) with 32 channels, with an output of \SI{\pm 10}{\volt}, a sampling time of \SI{390}{\nano\second} and a maximum output slew rate of \SI{20}{\volt\per\micro\second}. Its output is further amplified by a home-built, low-noise voltage amplifier with a voltage gain of \num{2.5} and a lower maximum slew rate of \SI{1}{\volt\per\micro\second}.
Finally, we filter the voltages to remove high-frequency noise using two sets of first-order low-pass filters, one at the output of the amplifier and one directly mounted in the cryostat on the trap carrier PCB, with a total cutoff frequency of \SI{30}{\kilo\hertz}.
The filters affect the AWG output, effectively limiting the maximum transport speed. Fixing the transport distance to \SI{375}{\micro\meter} (the space between Zone 1 and Zone 2) and the transport trajectory, filter effects begin to be relevant for waveforms whose total time approaches \SI{100}{\micro\second} or less, with a minimum time of \SI{\sim 25}{\micro\second} set by the filter time delay. Waveforms approaching this limit suffer from strong distortion of the voltages, with consequent deformation of the trapping potential which doesn't follow the required trajectory, potentially resulting in strong ion heating. To avoid this, we implement in-software filter precompensation and use it for the generation of waveforms of duration $< \SI{100}{\micro\second}$.

\end{appendix}

\bibliography{bibliography}

\end{document}